\documentclass[journal,twocolumn,10pt]{IEEEtran}

\usepackage{amsmath,cite,amsfonts,amssymb,color, mathtools,setspace,comment,float}
\usepackage{multicol}
\usepackage{subfigure}
\usepackage{graphicx}

\usepackage{multirow}
\usepackage{bigstrut}

\begin{document}

\title{Coordinated Regularized Zero-Forcing Precoding for Multicell MISO Systems with Limited Feedback}
\author{Jawad~Mirza,~\IEEEmembership{Student Member,~IEEE,}
        Peter~J.~Smith,~\IEEEmembership{Fellow,~IEEE,}
        and~Pawel~A.~Dmochowski,~\IEEEmembership{Senior Member,~IEEE}
        }

\maketitle

\begin{abstract}
We investigate coordinated regularized zero-forcing precoding for limited feedback multicell multiuser multiple-input single-output systems. We begin by deriving an approximation to the expected signal-to-interference-plus-noise ratio for the proposed scheme with perfect channel direction information (CDI) at the base station (BS). We also derive an expected SINR approximation for limited feedback systems with random vector quantization (RVQ) based codebook CDI at the BS. Using the expected interference result for the RVQ based limited feedback CDI, we propose an adaptive feedback bit allocation strategy to minimize the expected interference by partitioning the total number of bits between the serving and out-of-cell interfering channels. Numerical results show that the proposed adaptive feedback bit allocation method offers a spectral efficiency gain over the existing coordinated zero-forcing scheme.
\end{abstract}
\IEEEpeerreviewmaketitle

\section{Introduction}\label{intro}

\IEEEPARstart{I}{n} multicell systems, due to neighboring co-channel cells, the level of interference is high, especially at the cell-edge, thus degrading the spectral efficiency of the cell. Such a loss can be mitigated using BS coordination, where information is exchanged among the BSs via a backhaul link to suppress the inter-cell interference (ICI) in the downlink \cite{4487516}. 

In codebook-based limited feedback multiuser (MU) multiple-input multiple-output (MIMO) systems \cite{jindal2006mimo}, the user feeds back the index of the appropriate codebook entry or codeword to the BS, via a low-rate feedback link. This information is then used to compute precoders for the users. 
In \cite{bhagavatula2011adaptive}, a limited feedback strategy for MISO multicell systems at high signal-to-noise ratio (SNR) is developed using random vector quantization (RVQ) codebooks \cite{au2007performance}. 
An adaptive bit allocation method which maximizes the spectral efficiency is proposed in \cite{zhang2010adaptive} for limited feedback systems.
In \cite{5648782}, an adaptive feedback scheme for limited feedback MISO systems is proposed with a zero-forcing (ZF) precoding scheme which minimizes the expected spectral efficiency loss. 

Regularized zero-forcing (RZF) \cite{1391204} is a linear precoding technique shown to be effective for single-cell communication systems. RZF has also been extensively used in the analysis of 5G technologies such as massive MIMO \cite{HoydisBD13}.
Despite the numerous studies on coordinated multicell systems, little attention has been paid to coordinated RZF precoding prior to the development of massive MIMO \cite{6779600}. Thus, in this paper we investigate coordinated RZF precoding for conventional (small-scale) multicell MU MISO systems, where BSs share out-of-cell interfering CSI to coordinate transmission.

We also derive expected SINR approximations for the proposed scheme with perfect channel direction information (CDI) and with RVQ codebook CDI at the BS. Furthermore, we develop an adaptive bit allocation scheme that distributes the bits to serving and out-of-cell interfering channels, minimizing interference at users. We assume perfect knowledge of channel quality indicator (CQI) at the BS \cite{5648782}. 
The main contributions of this paper are summarized below.
\begin{itemize}
\item We investigate a coordinated RZF precoding scheme for multicell MU MISO systems, where interfering channels are shared among BSs.
\item Analytical expressions are derived to approximate the expected SINR for the proposed system with perfect CDI and limited feedback RVQ CDI.
\item We propose a novel adaptive bit allocation method that minimizes ICI.
\end{itemize}
\section{Downlink System Model}
\begin{figure*}[!ht]
\normalsize
\setcounter{equation}{2}
\begin{equation}\label{2}
 \textrm{SINR}_{l,k} = \frac{ \frac{P_{l,k,k}}{\gamma_{k}} \left| \mathbf{h}_{l,k,k} \mathbf{w}_{l,k} \right|^2}{1 +\frac{P_{l,k,k}}{\gamma_{k}} \sum_{\substack{m=1\\m \neq l}}^{L} \left| \mathbf{h}_{l,k,k} \mathbf{w}_{m,k} \right|^2 + \sum_{\substack{j=1\\j \neq k}}^{K} \frac{P_{l,k,j}}{\gamma_{j}} \sum_{q=1}^{L} \left| \mathbf{h}_{l,k,j} \mathbf{w}_{q,j}\right|^2}.
 \end{equation}
 \begin{equation}\label{3}
 \mathbb{E}\left[\textrm{SINR}_{l,k}\right] \approx \frac{\frac{P_{l,k,k}}{\bar{\gamma}_k} \mathbb{E}\left[\left| \mathbf{h}_{l,k,k} \mathbf{w}_{l,k} \right|^2\right]}{1+\frac{P_{l,k,k}}{\bar{\gamma}_k} \sum_{\substack{m=1\\m \neq l}}^{L} \mathbb{E}\left[\left| \mathbf{h}_{l,k,k} \mathbf{w}_{m,k} \right|^2\right] + \sum_{\substack{j=1\\j \neq k}}^{K} \frac{P_{l,k,j}}{\bar{\gamma}_j} \sum_{q=1}^{L} \mathbb{E} \left[\left| \mathbf{h}_{l,k,j} \mathbf{w}_{q,j}\right|^2\right]},
 \end{equation}
 \hrulefill
\vspace*{-1em}
\end{figure*}
\setcounter{equation}{0}
Consider a multicell  MU MISO system with $K$ cells having a single BS each. Each BS has $M$ transmit antennas and simultaneously serves $L$ single antenna users with $KL\leq M$\footnote{We assume $KL < M$, as at high SNR, RZF is equivalent to ZF.}. All the $K$ cells are interconnected via backhaul links assumed to be error free without delay. The $1 \times M$ channel vector between the $l^{\textrm{th}}$ user in the $k^{\textrm{th}}$ cell and the serving BS is given by $\mathbf{h}_{l,k,k}$. The interfering channel vector between the $l^{\textrm{th}}$ user in the $k^{\textrm{th}}$ cell and the $j^{\textrm{th}}$ interfering BS is denoted by $\mathbf{h}_{l,k,j}$, where $j \neq k$. The channel entries $\mathbf{h}_{l,k,k}$ and $\mathbf{h}_{l,k,j}$ are independent and identically distributed (i.i.d.) complex Gaussian $\mathcal{CN}(0,1)$. The downlink received signal at the $l^{\textrm{th}}$ user in the $k^{\textrm{th}}$ cell is given by\footnote{$(\cdot)^H$, $(\cdot)^T$ and $(\cdot)^{-1}$ denote the conjugate transpose, the transpose and the inverse operations, respectively. $\|\cdot\|$ and $|\cdot|$ stand for vector and scalar norms, respectively. $\mathbb{E}[\cdot]$ denotes statistical expectation.}
\begin{align}
 y_{l,k} \hspace{-.3em}&= \hspace{-.3em} \sqrt{\frac{P_{l,k,k}}{\gamma_{k}}} \mathbf{h}_{l,k,k} \mathbf{w}_{l,k} s_{l,k} \hspace{-.1em}+ \hspace{-.1em} \sqrt{\frac{P_{l,k,k}}{\gamma_{k}}} \hspace{-.2em}\sum_{\substack{m=1\\ m\neq l}}^{L} \hspace{-.2em}\mathbf{h}_{l,k,k} \mathbf{w}_{m,k} s_{m,k}\nonumber \\
 &+ \sum_{j=1, j\neq k}^{K} \sqrt{\frac{P_{l,k,j}}{\gamma_{j}}} \mathbf{h}_{l,k,j} \sum_{q=1}^{L} \mathbf{w}_{q,j} s_{q,j} + n_{l,k},\label{rec}
\end{align}
where $\mathbf{w}_{l,k}$ is the non-normalized precoding vector for the $l^{\textrm{th}}$ user in the $k^{\textrm{th}}$ cell and $\gamma_k$ is the normalization parameter (to be discussed later) for the  $k^{\textrm{th}}$ cell. $s_{l,k}$ and $n_{l,k}\sim\mathcal{CN}(0,N_0)$ denote the data symbol and the noise for the $l^{\textrm{th}}$ user in the $k^{\textrm{th}}$ cell. The data symbols are selected from the same constellation with $\mathbb{E} \left[ | s_{l,k} |^2 \right] = 1$. $P_{l,k,k}$ and $P_{l,k,j}$ are the received powers at the $l^{\textrm{th}}$ user in the $k^{\textrm{th}}$ cell from serving and interfering BSs, respectively, given by 
 \begin{figure}[!t]
  \centering
 \includegraphics[width=7cm, height=5cm]{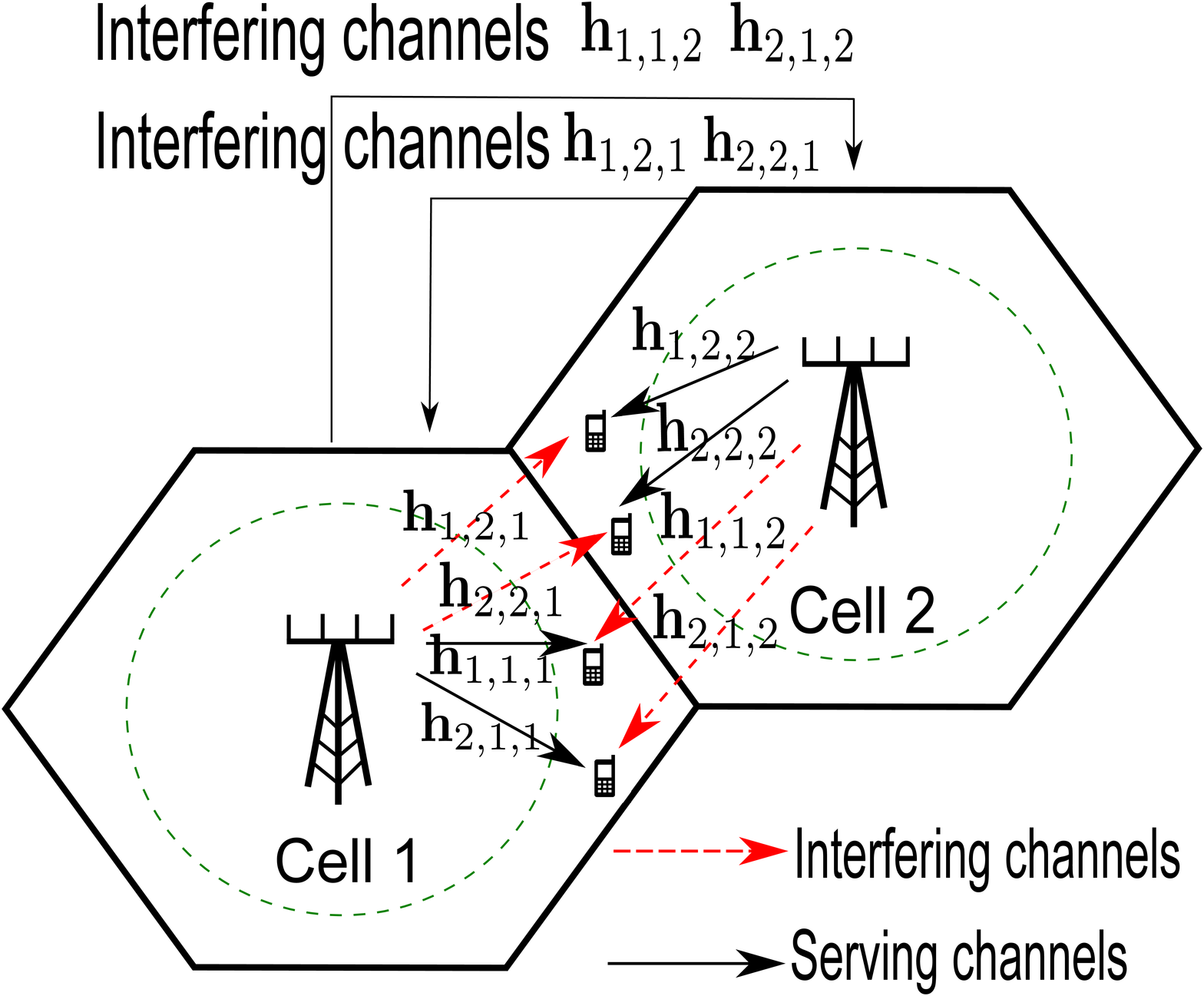}
  \caption{The system model for $K=2$ and $L=2$ cell-edge users.}
  \label{Fig1}
\end{figure}
\begin{equation}\label{p_loss}
 P_{l,k,k}=P_0  \left(\frac{R}{d_{l,k,k}}\right)^a \hspace{-.2em}z_{l,k,k}, \ \ \  P_{l,k,j}=P_0 \left(\frac{R}{d_{l,k,j}}\right)^a \hspace{-.2em}z_{l,k,j} \nonumber
\end{equation}
where $P_0$ is the power received at the distance $R$ in the absence of shadowing, $R$ is the cell radius and $a$ is the path loss exponent. The shadowing is modeled as a log-normal random variable, $z_{l,k,k}=10^{(\eta_{l,k,k}\sigma_{SF}/10)}$, for the channel between the $l^{\textrm{th}}$ user in the $k^{\textrm{th}}$ cell and the $k^{\textrm{th}}$ BS, where $\sigma_{SF}$ is the shadowing standard deviation in dB and $\eta_{l,k,k}\sim \mathcal{CN}(0,1)$. $d_{l,k,k}$ and $d_{l,k,j}$ are the distances from the serving BS and the interfering BS to the $l^{\textrm{th}}$ user in the $k^{\textrm{th}}$ cell, respectively.

\section{Coordinated RZF precoding}\label{perff}
In this study, the serving BS applies RZF not only to the channels of the same cell users but also considers its interfering channels to the users located in the adjacent cells, thus mitigating or suppressing the interference it causes to those users. All interfering channels and the serving channel are determined by the user using cell-specific pilots and these channels are conveyed to the serving BS. The interfering channels caused by the respective BSs are delivered to them via backhaul links. 
The system model for $K=2$ cells and $L=2$ cell-edge users with serving and interfering channels is shown in Fig.~\ref{Fig1}.
The non-normalized RZF precoder ${\mathbf{w}}_{l,k}$, for the $l^{\textrm{th}}$ user in the $k^{\textrm{th}}$ cell is the  $l^{\textrm{th}}$ column of\footnote{Like \cite{6678041,6779600}, path loss and shadowing are not considered in \eqref{1}.} \cite{1391204} 

\setcounter{equation}{8}
\begin{figure*}[!ht]
\normalsize
\begin{flalign}
 \text{F}_k &=   \mathbb{E}\left[\left(\sum_{n=1}^M \frac{\lambda_n}{\lambda_n + \alpha_k}\right)^2 \right] = \text{D}_k^{(2)} + \sum_{i=1}^M \sum_{j=1,j\neq i}^{M} && \nonumber \\
 &  \left( \sum_{r=0}^{i-1} \sum_{s=0}^{i-1} \left( -1 \right)^{r+s} {i-1 \choose i-1-r}{i-1 \choose i-1-s} \frac{1}{r! \ s!} \sum_{b=0}^{1+r+s} {1+r+s \choose b} (-\alpha_k)^{1+r+s-b} \textrm{e}^{\alpha_k}
  \int_{\alpha_k}^{\infty} v^{b-1} \textrm{e}^{-v} dv \right)^2 -&& \nonumber \\
  & \left( \sum_{r=0}^{i-1} \sum_{s=0}^{j-1} \left( -1 \right)^{r+s} {i-1 \choose i-1-r}{j-1 \choose j-1-s} \frac{1}{r! \ s!} \sum_{b=0}^{1+r+s} {1+r+s \choose b} (-\alpha_k)^{1+r+s-b} \textrm{e}^{\alpha_k}   \int_{\alpha_k}^{\infty} v^{b-1} \textrm{e}^{-v} dv \right)^2&& \label{n_eq1} 
\end{flalign}
\hrulefill
\vspace*{-1em}
\end{figure*}
\setcounter{equation}{1}
\begin{equation}\label{1}
 \mathbf{W}_k = \mathbf{H}_k^{H}\left(\mathbf{H}_k\mathbf{H}_k^H + \alpha_k \mathbf{I}\right)^{-1},
\end{equation}
where $\mathbf{H}_k=  [ \mathbf{X}_1^T \ \mathbf{X}_2^T \ldots \mathbf{X}_k^T \ldots \mathbf{X}_K^T ]^T$ is a $KL \times M$ concatenated matrix, with $\mathbf{X}_1= [\mathbf{h}_{1,1,k}^T \ldots \mathbf{h}_{L,1,k}^T ]^T$. 
The resulting precoding matrix is normalized, such that $\tilde{\mathbf{W}}_k = \mathbf{W}_k/\sqrt{\gamma_k}$, where $\gamma_k=\|\mathbf{W}_k\|_F^2/M$, to satisfy the total power constraint. The regularization parameter for the $k^\textrm{th}$ BS is denoted by $\alpha_k$ (discussed in Section \ref{rp}).  When perfect CDI and CQI are available at the BS\footnote{The CDI and CQI for the channel between the $l^{\textrm{th}}$ user in the $k^{\textrm{th}}$ cell and the $k^{\textrm{th}}$ BS, are defined as, CDI$=\mathbf{h}_{l,k,k}/\| \mathbf{h}_{l,k,k}\|$ and CQI $=\| \mathbf{h}_{l,k,k}\|$.}, the SINR expression for the $l^{\textrm{th}}$ user in the $k^{\textrm{th}}$ cell can be written as \eqref{2} \cite{6678041,6779600}. We can take the expectation of the SINR in \eqref{2} and approximate it as \eqref{3} using the expected SINR approximation in \cite{5510182}, 
where $\bar{\gamma}_k = \mathbb{E}\left[ \gamma_k \right]$ and $\bar{\gamma}_j = \mathbb{E}\left[ \gamma_j \right]$. Note that the numerator and  the denominator are dependent as they share some random variables in common. This complicates the calculation of the mean SINR and we therefore employ the SINR approximation approach given in \cite{5510182}. This approximation has been shown to get tighter as $M$ grows large. We evaluate \eqref{3} by computing the expected terms for the $KL=M$ case.
\setcounter{equation}{4}

\textbf{Expected signal power:} The expected signal power in \eqref{3} is
\begin{equation}\label{i3}
 S_{l,k}= \frac{P_{l,k,k}}{\bar{\gamma}_{k}} \mathbb{E} \left[\left| \mathbf{h}_{l,k,k} \mathbf{w}_{l,k} \right|^2\right].
\end{equation}
Using the eigenvalue decomposition, $\mathbf{H}_k\mathbf{H}_k^H=\mathbf{Q}\mathbf{\Lambda}\mathbf{Q}^H$, the expectation in \eqref{i3}, denoted by $\delta_{l,k}$, is written as \cite{1391204}
\begin{flalign}
\delta_{l,k}=\mathbb{E} \left[\left| \mathbf{h}_{l,k,k} \mathbf{w}_{l,k} \right|^2 \right] &=\mathbb{E} \left[ \left(\sum_{n=1}^{M} \frac{\lambda_n}{\lambda_n + \alpha_{k}} |q_{l,n}|^2\right)^2 \right], \nonumber 
\end{flalign}
where $\lambda_n$ is the $n^{\textrm{th}}$ eigenvalue corresponding to the $n^{\textrm{th}}$ diagonal entry of $\mathbf{\Lambda}$ and $q_{l,n}$ denotes the entry of $\mathbf{Q}$ corresponding to the $l^{\textrm{th}}$ row and $n^{\textrm{th}}$ column. 
Using \cite{1391204}, the expectation over $\mathbf{Q}$ yields
\begin{align}
\delta_{l,k} =&\frac{1}{v}\mathbb{E}_{\lambda} \left[\left(\sum_{n=1}^{M} \frac{\lambda_n}{\lambda_n + \alpha_{k}}\right)^2\right] 
+ \frac{1}{v} \mathbb{E}_{\lambda} \left[ \sum_{n=1}^{M} \left(\frac{\lambda_n}{\lambda_n + \alpha_{k}}\right)^2 \right]\label{5}
\end{align}
where $v=1/M(M+1)$. The value of $\bar{\gamma}_k$ is given by
\begin{flalign}
 \bar{\gamma}_k &=  \frac{1}{M}\mathbb{E} \left[ \| \mathbf{W}_k\|_F^2\right]
  =\frac{1}{M} \mathbb{E}_{\lambda} \left[\sum_{n=1}^{M} \frac{\lambda_n}{\left(\lambda_n + \alpha_{k}\right)^2}\right].  \label{i5}
\end{flalign}
Expectations in \eqref{5} and \eqref{i5} are solved in Result 1 and 2.

\emph{\textbf{Result 1:} When the entries of an $M \times M$ matrix $\mathbf{H}$ are i.i.d. $\mathcal{CN}(0,1)$ random variables, the expected value of $\sum_{n=1}^M \frac{\left(\lambda_n\right)^t}{\left(\lambda_n + \alpha_{k}\right)^2} $, where $\lambda_n$ is the $n^{\textrm{th}}$ eigenvalue of $\mathbf{H}\mathbf{H}^H$ with respect to $\lambda_n$ $\forall n$, is given by}
\begin{flalign}
 \text{D}_k^{(t)} &=  \mathbb{E}\left[\sum_{n=1}^M \frac{\left(\lambda_n\right)^t}{\left(\lambda_n + \alpha_{k} \right)^2}\right] \label{fc6}&&  \\
 =& \sum_{i=1}^M \sum_{j=0}^{i-1} \sum_{l=0}^{i-1} \left( -1 \right)^{j+l} {i-1 \choose i-1-j}{i-1 \choose i-1-l} \frac{1}{j! \ l!} && \nonumber \\
 &\sum_{s=0}^{t+j+l} {t+j+l \choose s} (-\alpha_k)^{t+j+l-s} \textrm{e}^{\alpha_k} \int_{\alpha_k}^{\infty} v^{s-2} \textrm{e}^{-v} dv.\nonumber &&
\end{flalign}

\emph{Proof:} See Appendix \ref{aa1}.

\emph{\textbf{Result 2:} When the entries of an $M \times M$ matrix $\mathbf{H}$ are i.i.d. $\mathcal{CN}(0,1)$ random variables, then the expected value of $\left(\sum_{n=1}^M \frac{\lambda_n}{\lambda_n + \alpha_{k}}\right)^2 $, where $\lambda_n$ is the $n^{\textrm{th}}$ eigenvalue of $\mathbf{H}\mathbf{H}^H$ with respect to $\lambda_n$ $\forall n$, is given by \eqref{n_eq1}.}
\setcounter{equation}{9}

\noindent\emph{Proof:} See Appendix \ref{a2}. 

Using Result 1 and 2, we can write \eqref{5} and \eqref{i5} as
\begin{equation}\label{ssi}
 \delta_{l,k} = \frac{\text{F}_k + \text{D}_k^{(2)}}{M\left(M+1\right)}
\end{equation}
\begin{equation}\label{ssi5}
  \bar{\gamma}_k   =\frac{ \text{D}_k^{(1)}}{M}.
\end{equation}
Therefore, the expected signal power \eqref{i3} becomes
\begin{equation}
 S_{l,k} =  \frac{P_{l,k,k}}{\bar{\gamma}_{k}} \delta_{l,k}.
\end{equation}
\textbf{Expected interference power:} The expected interference power in \eqref{3} is
\begin{align}
 I_{l,k} &= \frac{P_{l,k,k}}{\bar{\gamma}_k} \sum_{m=1, m \neq l}^{L} \mathbb{E}\left[\left| \mathbf{h}_{l,k,k} \mathbf{w}_{m,k} \right|^2\right] \nonumber \\
 &+ \sum_{j=1, j \neq k}^{K} \frac{P_{l,k,j}}{\bar{\gamma}_j} \sum_{q=1}^{L} \mathbb{E} \left[\left| \mathbf{h}_{l,k,j} \mathbf{w}_{q,j}\right|^2\right].\label{intpow}
\end{align}
In order to evaluate \eqref{intpow}, we observe that
\begin{equation}
\mathbb{E}\left[\left| \mathbf{h}_{l,k,k} \mathbf{w}_{m,k} \right|^2\right] = \mathbb{E} \left[\left| \mathbf{h}_{l,k,j} \mathbf{w}_{q,j}\right|^2\right] \overset{\Delta}{=} \psi  .
\end{equation}
Hence,
\begin{equation}
 I_{l,k} = \frac{P_{l,k,k}}{\bar{\gamma}_k} (L-1) \psi + \sum_{j=1, j \neq k}^{K} \frac{P_{l,k,j}}{\bar{\gamma}_j}  L \psi.
\end{equation}
Now $\psi$ can be found from $\varrho/(M-1)$ where
\begin{equation}\label{psi_sum}
\varrho \hspace{-.2em}= \hspace{-.5em}\sum_{\substack{m=1\\m \neq l}}^{L} \hspace{-.2em}\mathbb{E}\left[\left| \mathbf{h}_{l,k,k} \mathbf{w}_{m,k} \right|^2\right]\hspace{-.2em} + \hspace{-.2em}\sum_{\substack{j=1\\j \neq k}}^{K} \sum_{q=1}^{L} \mathbb{E} \left[\left| \mathbf{h}_{l,k,j} \mathbf{w}_{q,j}\right|^2\right]
\end{equation}
We note that \eqref{psi_sum} is the expected interference and is the difference between the expected total received power and the expected signal power \cite{1391204}. Hence,  \eqref{psi_sum} becomes
\begin{align}
 \varrho &= \xi_k - \delta_{l,k}  =  \frac{\text{D}_k^{(2)}}{M} - \frac{\text{F}_k+\text{D}_k^{(2)}}{M\left( M+1\right)},\label{psii}
\end{align}
where $\xi_k$ is the expected total received signal given by
$\xi_k = \mathbb{E} \left[\| \mathbf{H}_k \mathbf{W}_k\|_F^2 \right]/M = \text{D}_k^{(2)}/M$.

\textbf{Expected SINR with perfect CDI:} We can write the expected SINR in \eqref{3} in terms of $\delta_{l,k}$, $\psi$, $\bar{\gamma}_k$ and $\bar{\gamma}_j$ as
\begin{equation}\label{6}
 \mathbb{E}\left[\textrm{SINR}_{l,k}\right] \hspace{-.2em} \approx \hspace{-.2em}\frac{\frac{P_{l,k,k}}{\bar{\gamma}_k} \delta_{l,k}}{1+ \frac{ (L-1)P_{l,k,k}}{\bar{\gamma}_k} \psi + \sum_{\substack{j=1\\j \neq k}}^{K} \frac{LP_{l,k,j}}{\bar{\gamma}_j}  \psi}.
\end{equation}
\setcounter{equation}{24}
\begin{figure*}[!ht]
\normalsize
\begin{align}
 \mathbb{E}  \left[\widetilde{\textrm{SINR}}_{l,k}\right] \hspace{-.3em}\approx\hspace{-.3em}
  \frac{\frac{P_{l,k,k}}{\bar{\gamma}_k} \mathbb{E} \left[\left|\left(\hat{\mathbf{h}}_{l,k,k} + {\mathbf{e}}_{l,k,k}\right) \hat{\mathbf{w}}_{l,k}\right|^2 \right]}{1 + \frac{P_{l,k,k}}{\bar{\gamma}_k} \sum_{\substack{m=1\\ m\neq l}}^{L} \mathbb{E}  \left[ \left|\left(\hat{\mathbf{h}}_{l,k,k} + {\mathbf{e}}_{l,k,k}\right) \hat{\mathbf{w}}_{m,k}\right|^2 \right] \hspace{-.3em}+ \sum_{\substack{j=1\\j\neq k}}^{K} \frac{P_{l,k,j}}{\bar{\gamma}_j} \sum_{q=1}^{L} \mathbb{E}  \left[\left|\left(\hat{\mathbf{h}}_{l,k,j} + {\mathbf{e}}_{l,k,j}\right)  \hat{\mathbf{w}}_{q,j}\right|^2 \right] } \label{rvqsinr}
\end{align}
\hrulefill
\vspace*{-1em}
\end{figure*}
\setcounter{equation}{18}
\textbf{Non-coordinated interference:} In the presence of out-of-cell non-coordinated interference, an additional interference term in the denominator of \eqref{6} is added, yielding
\begin{align}
 &\mathbb{E}\left[\textrm{SINR}_{l,k}\right] \approx  \label{61}\\
 &\frac{\frac{P_{l,k,k}}{\bar{\gamma}_k} \delta_{l,k}}{1+ \frac{P_{l,k,k}}{\bar{\gamma}_k} (L-1) \psi + \sum_{\substack{j=1\\j \neq k}}^{K} \frac{P_{l,k,j}}{\bar{\gamma}_j}  L \psi + \sum_{c=1}^C P_{l,k,c} \Upsilon}\nonumber ,
\end{align}
where $C$ is the total number of non-coordinated interfering cells in the system, $P_{l,k,c}$ is the received signal power at the $l^{\textrm{th}}$ user in the $k^{\textrm{th}}$ cell from the $c^{\textrm{th}}$ non-coordinated BS. The quantity $\Upsilon$ is given by
\begin{equation}
 \Upsilon = \mathbb{E} \left[ \sum_{q=1}^L \left| \mathbf{h}_{l,k,c} \tilde{\mathbf{w}}_{q,c} \right|^2 \right],\label{611}
\end{equation}
where $\mathbf{h}_{l,k,c}$ is the channel between the $l^{\textrm{th}}$ user in the $k^{\textrm{th}}$ cell and the $c^{\textrm{th}}$ non-coordinated BS. $\tilde{\mathbf{w}}_{q,c}$ denotes the normalized precoding vector for the $q^{\textrm{th}}$ user in the $c^{\textrm{th}}$ non-coordinated cell, given by $\tilde{\mathbf{w}}_{q,c}= \mathbf{w}_{q,c} / \sqrt{\gamma_c}$, where $\mathbf{w}_{q,c}$ is the non-normalized precoding vector for the $q^{\textrm{th}}$ user in the $c^{\textrm{th}}$ non-coordinated cell and $\gamma_c$ denotes the normalization parameter for the RZF precoding matrix at the $c^{\textrm{th}}$ non-coordinated BS. We can write
\begin{align}
 \Upsilon &= L \mathbb{E} \left[ \left| \mathbf{h}_{l,k,c} \tilde{\mathbf{w}}_{q,c} \right|^2 \right] = L \mathbb{E} \left[ \mathbf{h}_{l,k,c} \tilde{\mathbf{w}}_{q,c} \tilde{\mathbf{w}}_{q,c}^H \mathbf{h}_{l,k,c}^H \right] \nonumber \\
  &= L \mathbb{E} \left[ \textrm{Tr} (\tilde{\mathbf{w}}_{q,c} \tilde{\mathbf{w}}_{q,c}^H) \right] = L \mathbb{E} \left[ \tilde{\mathbf{w}}_{q,c}^H \tilde{\mathbf{w}}_{q,c} \right] = L,\label{6111}
\end{align}
where $\textrm{Tr}(\cdot)$ denotes the trace of the matrix and \eqref{6111} follows from the fact that $\mathbf{h}_{l,k,c}$ and $\tilde{\mathbf{w}}_{q,c}$ are independent vectors and $\| \tilde{\mathbf{w}}_{q,c} \|^2 $ has a unit mean.

\section{Limited Feedback with RVQ Codebooks}\label{rvvq}

In FDD communication systems, limited feedback techniques are often used to equip the BS with knowledge of the CSI. A common codebook is maintained at the BS and the user, such that the user feeds back the index of the appropriate codeword to the BS via a low-rate link. In this section, we study the impact of RVQ codebooks \cite{au2007performance} on the performance of coordinated RZF precoding.

The user quantizes the estimated CDI (here, perfect estimation is assumed) using a codebook. The quantized channel vector of the $l^{\textrm{th}}$ user in the $k^{\textrm{th}}$ cell is denoted by $\hat{\mathbf{h}}_{l,k,k}$.  We consider an RVQ codebook where $B_{\textrm{total}}$ is the total number of feedback bits at the user. Each user quantizes serving and out-of-cell interfering channels, thus we can write $B_{\textrm{total}}=\sum_{i=1}^{K} B_{l,k,i}$ where $B_{l,k,i}$ is the number of bits used to quantize the channel between the $l^{\textrm{th}}$ user in the $k^{\textrm{th}}$ cell and the $i^{\textrm{th}}$ BS. The perfect concatenated channel matrix for the $k^{\textrm{th}}$ BS is modeled as \cite{4686268}
\begin{equation}
\mathbf{H}_k=\hat{\mathbf{H}}_k + \mathbf{E}_k,
\end{equation}
where $\mathbf{E}_k = [ \mathbf{G}_1^T \ \mathbf{G}_2^T \ldots \mathbf{G}_k^T \ldots \mathbf{G}_K^T ]^T$ is a $KL \times M$ quantization error matrix, $\mathbf{G}_1= [\mathbf{e}_{1,1,k}^T \ldots \mathbf{e}_{L,1,k}^T ]^T$ and $\mathbf{H}_k \sim \mathcal{CN}(0,1)$. The quantization error vector between the $l^{\textrm{th}}$ user in the $k^{\textrm{th}}$ cell and the $k^{\textrm{th}}$ BS is denoted by $\mathbf{e}_{l,k,k}$, with $\mathbf{e}_{l,k,k}\sim \mathcal{CN}(0, \sigma^2_{l,k,k}\mathbf{I})$.  Using an upper bound on the quantization error for RVQ codebooks in terms of squared chordal distance given in \cite{jindal2006mimo}, we have, $\sigma_{l,k,k}^2 \leq 2^{\frac{-B_{l,k,k}}{M-1}}$. In our model, we assume the worst case scenario, where $\sigma^2_{l,k,k} = 2^{\frac{-B_{l,k,k}}{M-1}}$. Similarly, the  quantized concatenated channel matrix, $\hat{\mathbf{H}}_k = [ \tilde{\mathbf{G}}_1^T \ \tilde{\mathbf{G}}_2^T \ldots \tilde{\mathbf{G}}_k^T \ldots \tilde{\mathbf{G}}_K^T ]^T$, where $\hat{\mathbf{H}}_k$ is a $KL \times M$ concatenated quantized channel matrix with $\tilde{\mathbf{G}}_1= [\hat{\mathbf{h}}_{1,1,k}^T \ldots \hat{\mathbf{h}}_{L,1,k}^T ]^T$. The entries of $\hat{\mathbf{H}}_k$ are $\hat{\mathbf{h}}_{l,k,k}\sim \mathcal{CN}(0, (1-\sigma^2_{l,k,k})\mathbf{I})$. 
The non-normalized precoding vector of the $l^{\textrm{th}}$ user in the $k^{\textrm{th}}$ cell, $\hat{\mathbf{w}}_{l,k}$, is the $l^{\textrm{th}}$ column of the matrix $\hat{\mathbf{W}}_k$, given by
\begin{equation}\label{vn1}
 \hat{\mathbf{W}}_k = \tilde{\mathbf{H}}_k^{H}\left(\tilde{\mathbf{H}}_k \tilde{\mathbf{H}}_k^H + \alpha_k \mathbf{I}\right)^{-1},
\end{equation}
where $\tilde{\mathbf{H}}_k = [ \tilde{\mathbf{X}}_1^T \ \tilde{\mathbf{X}}_2^T \ldots \tilde{\mathbf{X}}_k^T \ldots \tilde{\mathbf{X}}_K^T ]$ is a $KL \times M$ concatenated matrix with $\tilde{\mathbf{X}}_k= [\tilde{\mathbf{h}}_{1,k,k}^T \ldots \tilde{\mathbf{h}}_{L,k,k}^T ]^T$, and $\tilde{\mathbf{h}}_{l,k,k} = \hat{\mathbf{h}}_{l,k,k}/\sqrt{1-\sigma_{l,k,k}^2}$ such that $\tilde{\mathbf{H}}_k \sim \mathcal{CN} (0,1)$.
To meet the total power constraint, the precoding matrix is normalized by the parameter, ${\gamma}_k$, such that, $\bar{\mathbf{W}}_k = \hat{\mathbf{W}}_k/\sqrt{{\gamma}_k}$, where ${\gamma}_k = \|\mathbf{\hat{W}}_k\|_F^2/M$.
The received signal for the $l^{\textrm{th}}$ user in the $k^{\textrm{th}}$ cell is 
\begin{align}
 \hat{y}_{l,k} &= \sqrt{P_{l,k,k}/\gamma_k} \left(\hat{\mathbf{h}}_{l,k,k} + {\mathbf{e}}_{l,k,k}\right) \hat{\mathbf{w}}_{l,k} s_{l,k} \label{recq} \\ &+\sqrt{P_{l,k,k}/\gamma_{k}} \sum_{m=1, m\neq l}^{L}  \left(\hat{\mathbf{h}}_{l,k,k} + {\mathbf{e}}_{l,k,k}\right) \hat{\mathbf{w}}_{m,k} s_{m,k} \nonumber \\
 + & \sum_{j=1, j\neq k}^{K} \sqrt{P_{l,k,j}/\gamma_{j}}  \left(\hat{\mathbf{h}}_{l,k,j} + {\mathbf{e}}_{l,k,j}\right) \sum_{q=1}^{L} \hat{\mathbf{w}}_{q,j} s_{q,j} + n_{l,k}.\nonumber
\end{align}
The expected SINR for codebooks can be approximated by \eqref{rvqsinr} where $\bar{\gamma}_k=\mathbb{E}\left[ \gamma_k\right]$ and $\bar{\gamma}_j=\mathbb{E}\left[ \gamma_j\right]$.
\setcounter{equation}{25}

\textbf{Expected signal power:} The expectation of the signal power at the $l^{\textrm{th}}$ user in the $k^{\textrm{th}}$ cell in \eqref{rvqsinr} is given by
\begin{equation}\label{rvqsig}
 S_{l,k}^{'} = \frac{P_{l,k,k}}{\bar{\gamma}_k} \mathbb{E} \left[\left|\left(\hat{\mathbf{h}}_{l,k,k} + {\mathbf{e}}_{l,k,k}\right) \hat{\mathbf{w}}_{l,k}\right|^2 \right].
\end{equation}
We can write
\begin{flalign}
\Omega &= \mathbb{E} \left[ \left| \left(\hat{\mathbf{h}}_{l,k,k}  + {\mathbf{e}}_{l,k,k}\right) \hat{\mathbf{w}}_{l,k} \right|^2 \right] && \label{sigpow}\\
& \overset{(a)}{=} \mathbb{E} \left[ \left|\hat{\mathbf{h}}_{l,k,k}\hat{\mathbf{w}}_{l,k}\right|^2 \right] + \mathbb{E} \left[ \left| \mathbf{e}_{l,k,k}\hat{\mathbf{w}}_{l,k} \right|^2 \right]&& \nonumber \\
& \overset{(b)}{=}\left( 1 - \sigma_{l,k}^2 \right)\mathbb{E} \left[ \left|\tilde{\mathbf{h}}_{l,k,k}\hat{\mathbf{w}}_{l,k}\right|^2 \right]
 + \mathbb{E} \left[ \left| \mathbf{e}_{l,k,k}\hat{\mathbf{w}}_{l,k} \right|^2 \right],&& \nonumber
\end{flalign}
where in $(a)$ the expected value of the cross product terms is zero and in $(b)$ we use $\hat{\mathbf{h}}_{l,k,k}=\sqrt{1 - \sigma_{l,k}^2}\tilde{\mathbf{h}}_{l,k,k}$. Denoting the eigenvalue values of $\tilde{\mathbf{H}}_k\tilde{\mathbf{H}}_k^H = \tilde{\mathbf{Q}} \tilde{\mathbf{\Lambda}} \tilde{\mathbf{Q}}^{H}$ by $\tilde{\lambda}_1, \tilde{\lambda}_2, \ldots \tilde{\lambda}_M$, we can express \eqref{sigpow} as
\begin{align}
 \Omega &= \left( 1 - \sigma_{l,k}^2\right) \delta_{l,k} + \mathbb{E} \left[ | \mathbf{e}_{l,k,k} |^2 \right] \mathbb{E} \left[ \|\hat{\mathbf{W}}_{k} \|^2_{F} \right] /M \nonumber \\
  &= \left( 1 - 2^{\frac{-B_{l,k,k}}{M-1}}\right) \delta_{l,k} + \left(2^{\frac{-B_{l,k,k}}{M-1}}\right) \bar{\gamma}_k,\label{lfsig}
\end{align}
where $\delta_{l,k}$ and $\bar{\gamma}_k$ are given by \eqref{ssi} and \eqref{ssi5}, respectively. 
Thus, the expected signal power in \eqref{rvqsig} is
\begin{equation}\label{sig_rvqp}
 S_{l,k}^{'} = \frac{P_{l,k,k}}{\bar{\gamma}_k} \left[ \left( 1 - 2^{\frac{-B_{l,k,k}}{M-1}} \right)\delta_{l,k} + 2^{\frac{-B_{l,k,k}}{M-1}} \bar{\gamma}_k\right].
\end{equation}
\textbf{Expected interference power:} The expected interference in \eqref{rvqsinr} is given by
\begin{align}
 I_{l,k}^{'}&= \frac{P_{l,k,k}}{\bar{\gamma}_k} \sum_{m=1, m\neq l}^{L} \mathbb{E}  \left[ \left|\left(\hat{\mathbf{h}}_{l,k,k} + {\mathbf{e}}_{l,k,k}\right) \hat{\mathbf{w}}_{m,k}\right|^2 \right] \nonumber \\
 +& \hspace{-.4em}\sum_{j=1, j\neq k}^{K} \frac{P_{l,k,j}}{\bar{\gamma}_j} \sum_{q=1}^{L} \mathbb{E}  \left[\left|\left(\hat{\mathbf{h}}_{l,k,j} + {\mathbf{e}}_{l,k,j}\right)  \hat{\mathbf{w}}_{q,j}\right|^2 \right].\label{expint}
\end{align}
Again, the expected interference can be written as
\begin{flalign}
\tilde{\psi}_{l,k}=&\underbrace{  \mathbb{E}\left[ \left|\hat{\mathbf{e}}_{l,k,k} \hat{\mathbf{W}}_k\right|^2_F\right]+\mathbb{E} \left[ \left| \hat{\mathbf{h}}_{l,k,k} \hat{\mathbf{W}}_k \right|^2_F \right]}_{\textrm{total power at the user}}  \nonumber \\
&- \left[\left( 1 - 2^{\frac{-B_{l,k,k}}{M-1}} \right)\delta_{l,k} + \left(2^{\frac{-B_{l,k,k}}{M-1}}\right) \bar{\gamma}_k\right] \nonumber \\
=&   \bar{\gamma}_k M 2^{\frac{-B_{l,k,k}}{M-1}} + \left( 1 - 2^{\frac{-B_{l,k,k}}{M-1}} \right)\xi_k \nonumber \\
&- \left( 1 - 2^{\frac{-B_{l,k,k}}{M-1}} \right)\delta_{l,k} - \left(2^{\frac{-B_{l,k,k}}{M-1}}\right) \bar{\gamma}_k, \label{psi_k}
\end{flalign}
where $\xi_k=\frac{ \text{D}^{(2)}_k}{M}$. The interference at the $l^{\textrm{th}}$ user in the $k^{\textrm{th}}$ cell from the $j^{\textrm{th}}$ cell is
\begin{align}
 \tilde{\psi}_{l,j}& = \bar{\gamma}_j M 2^{\frac{-B_{l,k,j}}{M-1}} + \left( 1 - 2^{\frac{-B_{l,k,j}}{M-1}} \right)\xi_j \nonumber \\
  &- \left( 1 - 2^{\frac{-B_{l,k,j}}{M-1}} \right)\delta_{l,j} - \left(2^{\frac{-B_{l,k,j}}{M-1}}\right) \bar{\gamma}_j.\label{psi_j}
\end{align}
The interference from any single interfering source coming from $k^{\textrm{th}}$ and $j^{\textrm{th}}$ cell are $\psi_{l,k}^{'}=\tilde{\psi}_{l,k}/(M-1)$ and $\psi_{l,j}^{'}=\tilde{\psi}_{l,j}/(M-1)$, respectively and \eqref{expint} becomes
\begin{equation}\label{rvq_intt}
 I_{l,k}^{'}= \frac{P_{l,k,k}}{\bar{\gamma}_k} (L-1) \psi_{l,k}^{'} + \sum_{j=1, j\neq k}^{K} \frac{P_{l,k,j}}{\bar{\gamma}_j} L \psi_{l,j}^{'}.
\end{equation}
\textbf{Expected SINR with RVQ:} We can now express the expected SINR in \eqref{rvqsinr} using \eqref{sig_rvqp} and \eqref{rvq_intt}, as
\begin{align}
 &\mathbb{E} \left[\widetilde{\textrm{SINR}}_{l,k}\right] \approx \frac{\frac{P_{l,k,k}}{\bar{\gamma}_k} \left[\left( 1 - 2^{\frac{-B_{l,k,k}}{M-1}} \right)\delta_{l,k} + 2^{\frac{-B_{l,k,k}}{M-1}} \bar{\gamma}_k\right]}{ 1 + \frac{(L-1)P_{l,k,k}}{\bar{\gamma}_k}   \psi_{l,k}^{'} + \sum_{\substack{j=1\\j\neq k}}^{K} \frac{L P_{l,k,j}}{\bar{\gamma}_j}  \psi_{l,j}^{'}}. \label{rs}
\end{align}

\setcounter{equation}{40}
\begin{figure*}[!ht]
\normalsize
\begin{equation}\label{opbit}
 B_{l,k,i}^{*} = \min \left\{ B_{\textrm{total}}, \left\lfloor  \frac{B_{\textrm{total}}}{K} + \left(M-1 \right) \log_2 \left( \frac{\left( P_{l,k,i} (L-1)(1-\Delta_k) \right)^{\frac{K-1}{K}}}{\left(\prod_{j=1, j \neq i}^K  P_{l,k,j} L (1-\Delta_j) \right)^{\frac{1}{K}}} \right) \right\rfloor^{+} \right\},
 \end{equation}
\hrulefill
\vspace*{-1em}
\setcounter{equation}{34}
\end{figure*}
As $B_{\textrm{total}}\to \infty$, \eqref{rs} approaches the expected SINR approximation with perfect CDI \eqref{6}.

\textbf{Non-coordinated interference:} As for the perfect CDI case in Section \ref{perff}, we can also extend the expected SINR approximation for limited feedback systems in the presence of non-coordinated interfering cells, such that
\begin{align}
 &\mathbb{E} \left[\widetilde{\textrm{SINR}}_{l,k}\right] \approx \label{rs2} \\
 &\frac{\frac{P_{l,k,k}}{\bar{\gamma}_k} \left[\left( 1 - 2^{\frac{-B_{l,k,k}}{M-1}} \right)\delta_{l,k} + \left(2^{\frac{-B_{l,k,k}}{M-1}}\right) \bar{\gamma}_k\right]}{ 1 + \frac{(L-1)P_{l,k,k}}{\bar{\gamma}_k}   \psi_{l,k}^{'} + \sum_{\substack{j=1\\j\neq k}}^{K} \frac{LP_{l,k,j}}{\bar{\gamma}_j} \psi_{l,j}^{'}+ \sum_{c=1}^C P_{l,k,c} \Upsilon^{'}}, \nonumber
\end{align}
where $\Upsilon^{'}$ is defined as $\mathbb{E} \left[ \sum_{q=1}^L \left| \tilde{\mathbf{h}}_{l,k,c} \tilde{\mathbf{w}}_{q,c} \right|^2 \right]$ such that $\tilde{\mathbf{h}}_{l,k,c}=\hat{\mathbf{h}}_{l,k,c}+{\mathbf{e}}_{l,k,c}$ and $\tilde{\mathbf{w}}_{q,c}=\hat{\mathbf{w}}_{q,c}/\sqrt{\gamma_c}$, where $\gamma_c$ denotes the normalization parameter for the RZF precoding matrix at the $c^{\textrm{th}}$ non-coordinated BS. It is important to note that $\tilde{\mathbf{h}}_{l,k,c}$ and $\tilde{\mathbf{w}}_{q,c}$ are  independent and $\| \tilde{\mathbf{w}}_{q,c} \|^2$ has a unit mean, thus similarly to perfect CDI case, we have $\Upsilon^{'} = L$.

\section{Adaptive bit allocation method}\label{prpadp}
We now present adaptive feedback bit allocation with the proposed coordinated RZF scheme.
As discussed in Section \ref{intro}, there are numerous studies \cite{6397539,yu2012novel,5648782} on adaptive bit allocation for limited feedback systems. While, there are a few studies \cite{6585723}\cite{6779600} that consider RZF precoding with adaptive bit allocation for massive MISO systems, this problem is not well investigated for not so large antenna systems.

We propose an adaptive method to allocate the total number of bits at the user, $B_{\textrm{total}}=\sum_{i=1}^K B_{l,k,i}$, to quantize the serving channel and the out-of-cell interfering channels, by minimizing the mean interference at the user.  
The mean interference at the $l^{\textrm{th}}$ user in the $k^{\textrm{th}}$ cell in \eqref{rs}, is given by
\begin{align}
 \mathrm{I}_{l,k}^{'} &= \frac{P_{l,k,k}}{\bar{\gamma}^{'}_k} (L-1)  \psi_{l,k}^{'} + \sum_{j=1, j\neq k}^{K} \frac{P_{l,k,j}}{\bar{\gamma}^{'}_j} L \psi_{l,j}^{'} \label{int1} \\
 &=\frac{P_{l,k,k}}{\bar{\gamma}^{'}_k} \frac{(L-1) }{M-1} \tilde{\psi}_{l,k} + \sum_{j=1, j\neq k}^{K} \frac{P_{l,k,j}}{\bar{\gamma}^{'}_j} \frac{L}{M-1} \tilde{\psi}_{l,j}. \nonumber
\end{align}
Substituting the values of $\tilde{\psi}_{l,k}$ and $\tilde{\psi}_{l,j}$ from \eqref{psi_k} and \eqref{psi_j} into \eqref{int1} and rearranging, gives
\begin{flalign}\label{rrr}
I_{l,k}^{'} &=  P_{l,k,k} (L-1) 2^{\frac{-B_{l,k,k}}{M-1}} (1 - \Delta_k) +  P_{l,k,k} (L-1)\Delta_k&& \nonumber \\
+&\sum_{\substack{j=1\\j\neq k}}^{K}   P_{l,k,j}L 2^{\frac{-B_{l,k,j}}{M-1}}  (1- \Delta_j) \hspace{-.2em}+\hspace{-.3em}\sum_{\substack{j=1\\j\neq k}}^{K}   P_{l,k,j} L \Delta_j,&&
\end{flalign}
where $\Delta_k = (\xi_k - \delta_{l,k})/(\bar{\gamma}_k(M-1))$ and $\Delta_j = (\xi_j - \delta_{l,j})/(\bar{\gamma}_j(M-1))$. We can write \eqref{rrr} as
\begin{align}
I_{l,k}^{'} &= \underbrace{P_{l,k,k} (L-1)(1 - \Delta_k ) 2^{\frac{-B_{l,k,k}}{M-1}}}_{\bar{P}_{l,k,k}}\nonumber \\
&+ \sum_{j=1, j\neq k}^{K}   \underbrace{P_{l,k,j} L (1 - \Delta_j) 2^{\frac{-B_{l,k,j}}{M-1}}}_{\bar{P}_{l,k,j}} + P_I \label{rrr1}
\end{align}
where $P_I=P_{l,k,k} (L-1) \Delta_k   + \sum_{j=1, j\neq k}^{K}   P_{l,k,j} L \Delta_j$. In order to solve for the number of bits that minimizes the mean interference at the $l^{\textrm{th}}$ user in the $k^{\textrm{th}}$ cell, we define an optimization problem, given by
\begin{align}
  &\min_{B_{l,k,1}, \ldots, B_{l,k,K} \ \in \ \{0, \mathbb{R}^{+} \} } \ \sum_{i=1}^K \bar{P}_{l,k,i} 2^{\frac{-B_{l,k,i}}{M-1}}  \nonumber \\
 & s.t \ \ \sum_{i=1}^K B_{l,k,i} \leq B_{\textrm{total}}, \label{cff}
 \end{align}
where $\mathbb{R}^{+}$ denotes the set of positive real numbers. This is a convex optimization problem as the objective
function is logarithmically convex \cite{5648782}. We find the
solution in real space and discretize it to the nearest
integer \cite{5648782}. Using the Lagrangian function, we
define
\begin{equation}\label{op}
 L(B_{l,k,i},\lambda) = \sum_{i=1}^K \bar{P}_{l,k,i} 2^{\frac{-B_{l,k,i}}{M-1}} + \lambda \left( \sum_{i=1}^{K} B_{l,k,i} - B_{\textrm{total}}\right),
\end{equation}
where $\lambda$ denotes a Lagrange multiplier. Using the first order optimality Karush-Kuhn-Tucker (KKT) conditions we can solve \eqref{op} to obtain \eqref{opbit}, where $\lfloor \cdot \rfloor^{+} = \max \{ 0, \cdot \}$. 
Equation \eqref{opbit} yields the number of bits required by the $l^{\textrm{th}}$ user in the $k^{\textrm{th}}$ cell to quantize the serving and out-of-cell interfering channels, such that the mean interference is minimized.
\setcounter{equation}{41}
\section{regularization parameter analysis}\label{rp}
In \cite{nguyen2014mmse}, for single-cell non-homogeneous MU systems, the regularization parameter is defined as
\begin{equation}\label{aalph}
 \alpha_k = \frac{1}{L} \sum_{l=1}^{L} 1/P_{l,k,k},
\end{equation}
which we extend for a multicell system to
\begin{equation}\label{aalph2}
\tilde{\alpha}_k =  \frac{1}{KL} \sum_{c=1}^{K} \sum_{l=1}^{L} 1/P_{l,k,c}.
\end{equation}
In this study, we also consider an optimal regularization parameter, denoted by $\alpha_k^{opt}$, that maximizes the instantaneous spectral efficiency of the cell \cite{6585723}, given by
\begin{equation}\label{max_se}
 \alpha_k^{opt} = \arg \max_{\alpha_k^{opt} > 0} \sum_{l=1}^{L} \log_2 \left( 1 + \textrm{SINR}_{l,k}\right).
\end{equation}
Next, we numerically compute $\alpha_k^{opt} \in (0,\mathbb{R}^{+}) $, to evaluate the performance of the coordinated RZF scheme.
\section{Simulation Results}\label{sims}
\begin{figure}[!t]
  \centering
    \includegraphics[width=0.45\textwidth]{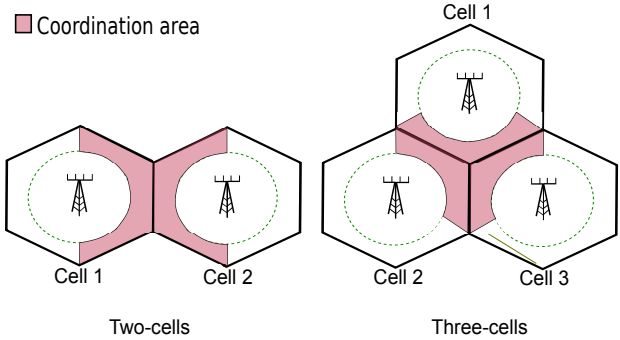}%
  \caption{Two-cell and three-cell coordination areas in the simulations.}
     \label{f1xx}
\end{figure}

We now present simulation results for the multicell MU MISO system with coordinated RZF precoding and compare it with non-coordinated RZF and coordinated ZF \cite{5648782} schemes. 
We use $R = 500$m, $\sigma_{SF}=8$dB and $a = 3.8$. We follow the coordination area definition in \cite{5648782}, i.e., coordination is needed when the user lies in the region $325\textrm{m} \leq d \leq 500\textrm{m}$, where $d$ is the distance of the user from the BS. The two- and three-cell coordination areas are illustrated in Fig. \ref{f1xx} where the cell-edge users are uniformly distributed. 
\subsection{SINR and spectral efficiency results}
We present SINR and spectral efficiency results for the proposed coordinated RZF scheme and plot these against the average received cell-edge SNR, $\rho_0$, given by $\rho_0 = \mathbb{E} \left[ 10\log_{10} \left(P_0 10^{(\eta\sigma_{SF}/10)}/N_0 \right) \right]$, with $N_0 =1$ and $\eta\sim\mathcal{N}(0,1)$.

\begin{figure}[!t]
   \centering
  \includegraphics[width=.5\textwidth]{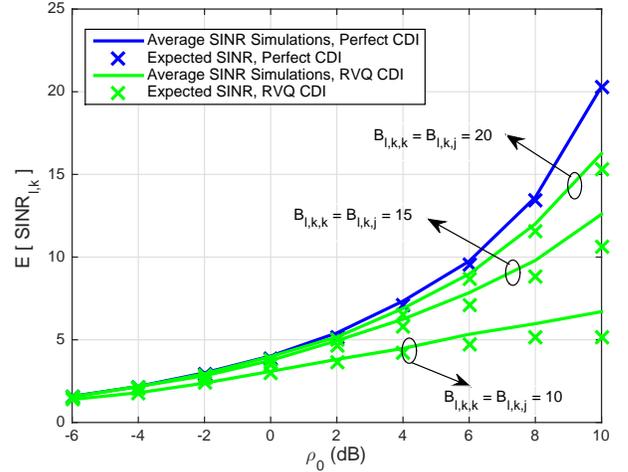}
  \caption{The average SINR of the user, with $K=2$, $C=0$, $L=2$, $M=4$ and $B_{l,k,k}=B_{l,k,j}=$ 20, 15 and 10, no shadowing.}
  \label{Fig000}
\end{figure}

Fig. \ref{Fig000} shows the average SINR performance of the coordinated RZF scheme with perfect CDI and limited feedback based RVQ CDI. The approximate expected SINRs derived in \eqref{6} and \eqref{rs} are plotted (in the linear scale) in Fig. \ref{Fig000}. It is observed that the approximations are tight, however the expected SINR approximation \eqref{rs} with smaller size RVQ codebooks shows a small deviation relative to the simulated average SINR at high $\rho_0$ values.
\begin{figure}[!t]
   \centering
  \includegraphics[width=.5\textwidth]{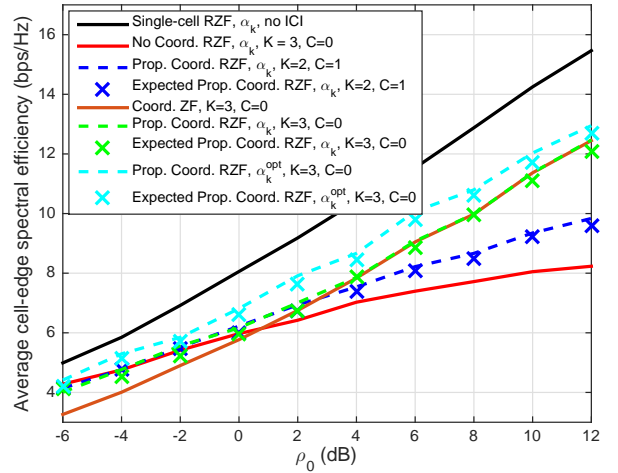}
  \caption{The average cell-edge spectral efficiency for $L=2$ and $M=8$.}
  \label{Figg2}
\end{figure}
The average cell-edge spectral efficiency for $L=2$ cell-edge users is shown in Fig.~\ref{Figg2} with $M=8$. Denoting the cell-edge spectral efficiency by $R_{\textrm{cell-edge}}$, its average is simulated by computing
\begin{equation}\label{errw}
 \mathbb{E} \left[ R_{\textrm{cell-edge}}\right]= \mathbb{E} \left[ \sum_{l=1}^L \log_2 \left( 1 + \textrm{SINR}_{l,k} \right)\right],
\end{equation}
where $\textrm{SINR}_{l,k}$ is given in \eqref{2} and the users are located in the cell-edge area. We refer to \eqref{errw} as the average cell-edge spectral efficiency of the cell. The single cell MU system gives superior average cell-edge spectral efficiency due to the absence of ICI. However, with ICI, the performance of the non-coordinated RZF precoding scheme suffers high losses at higher $\rho_0$ values. For the proposed coordinated RZF scheme, we consider two cases: 1) $K=2$ and $C=1$ and 2) $K=3$ and $C=0$. In Fig.~\ref{Figg2}, we consider two regularization parameters for the proposed coordinated RZF case 2: $\alpha_k$ and $\alpha_k^{opt}$.

The proposed coordinated RZF case 2 with $\alpha_k^{opt}$ achieves better average cell-edge spectral efficiency compared to the proposed coordinated RZF case 1 (with $\alpha_k$) and the non-coordinated RZF scheme. The proposed coordinated RZF schemes with both cases 1 and 2 outperform the coordinated ZF \cite{5648782} scheme. 
We also plot expected cell-edge spectral efficiency of the proposed scheme by using
\begin{equation}\label{errw1}
 \mathbb{E} \left[ \tilde{R}_{\textrm{cell-edge}}\right] \approx L  \log_2 \left( 1 + \mathbb{E} \left[\textrm{SINR}_{l,k}\right] \right),
\end{equation}
where $\mathbb{E} \left[\textrm{SINR}_{l,k}\right]$ is given in \eqref{6} and \eqref{61} for $C=0$ and $C\neq 0$ scenarios, respectively. It is seen that the approximations closely match the simulation results.

\begin{figure}[!t]
  \centering
  \subfigure[Co-channel interfering sectors.]{%
    \vspace{-5em}\includegraphics[width=.3\textwidth]{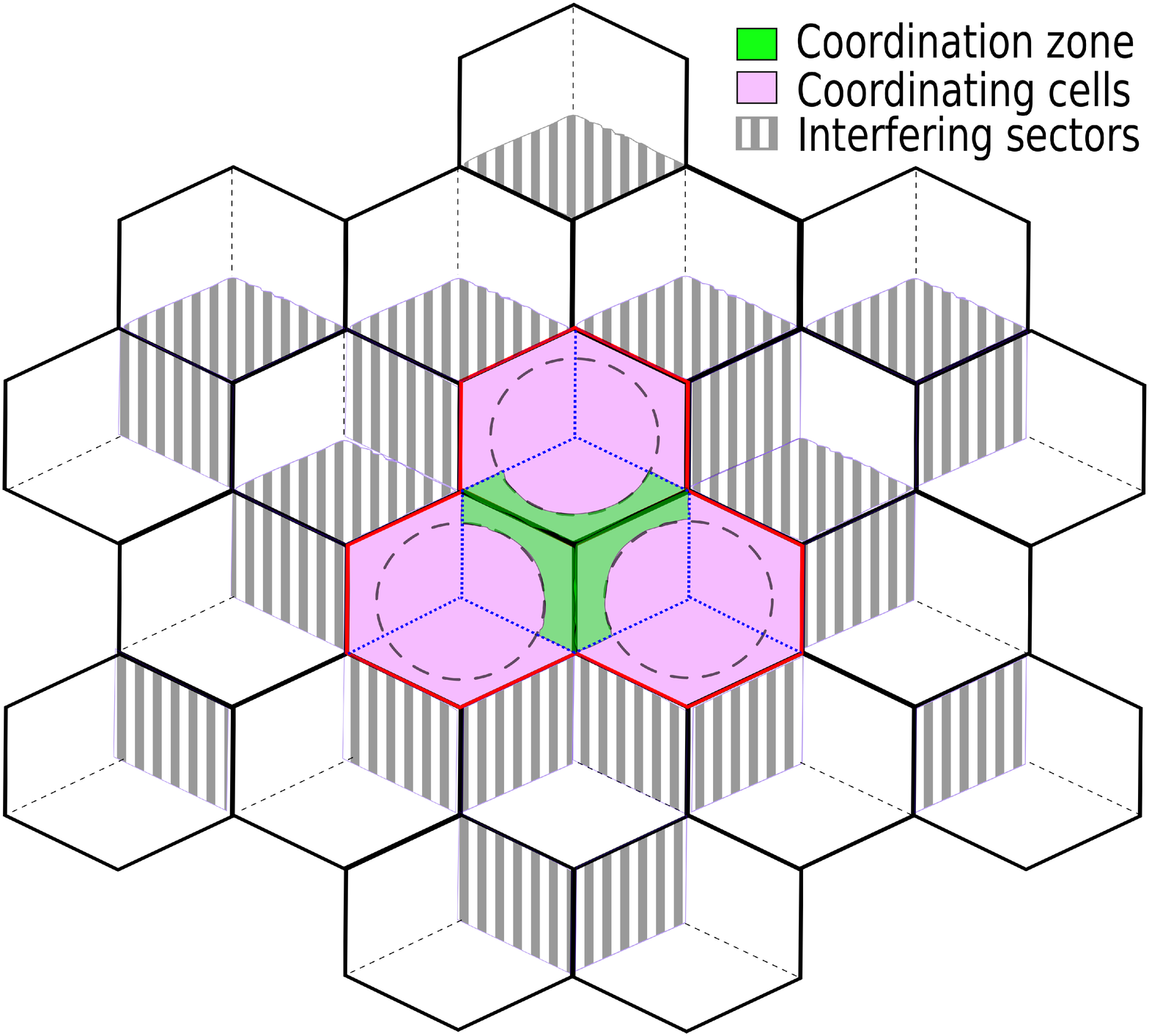}%
    \label{Fig2m}%
  }
   \subfigure[Average cell-edge spectral efficiency performance.]{%
    \includegraphics[width=0.5\textwidth]{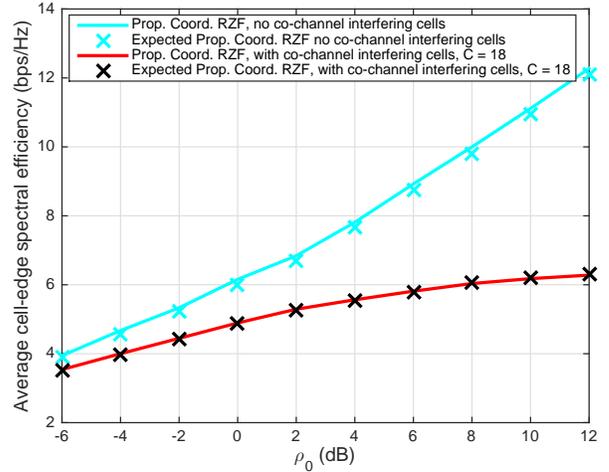}%
    \label{Figg2m}%
    }
  \caption{Cellular system with 21 cells and the average cell-edge spectral efficiency with $K=3$, $M=8$ and $L=2$.}
  \label{xxxxggm}
\end{figure}

In Fig. \ref{Fig2m}, we show 18 non-coordinated co-channel interfering cells surrounding the coordinated $K=3$ cells. Each cell consists of 3 sectors. For this scenario the average cell-edge spectral efficiency of the proposed coordinated RZF scheme is shown in Fig. \ref{Figg2m}. Each non-coordinated cell consists of $L=2$ users uniformly dropped in the cell, while the users in the $K=3$ coordinated cells are restricted to the coordination area near the cell-edge. The spectral efficiency performance in the absence of non-coordinated cells is obviously superior. On the other hand, with non-coordinated cells, the out-of-cell interference from the interfering sectors is high, thus massively reducing the performance of the system.  The expected cell-edge spectral efficiency approximation plotted using \eqref{errw1} with $\mathbb{E} \left[\textrm{SINR}_{l,k}\right]$ from \eqref{61}, matches closely with the simulations.
\begin{figure}[!t]
   \centering
  \includegraphics[width=.5\textwidth]{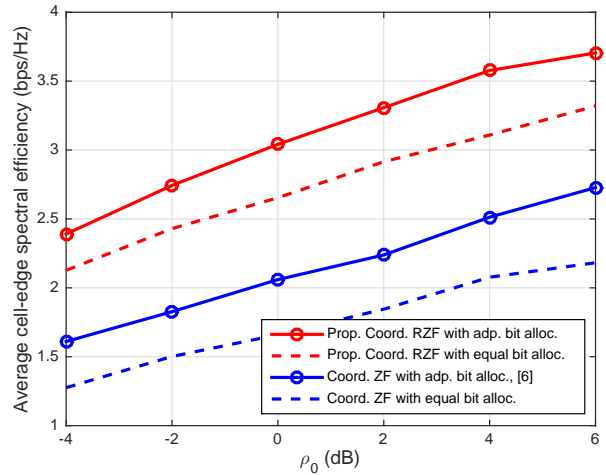}
  \caption{The average cell-edge spectral efficiency with $K=2$, $L=2$, $M=4$, $B_{\textrm{total}}=8$ and $\sigma_{SF}=8 $dB.}
  \label{Fig4}
\end{figure}
\subsection{Proposed adaptive bit allocation performance}
We evaluate the average cell-edge spectral efficiency of the proposed RZF precoding scheme with the adaptive bit allocation scheme discussed in Section \ref{prpadp}, with $K = 2 \ \textrm{and} \ 3$. From this point onwards, we use the regularization parameter $\tilde{\alpha}_k$ given in \eqref{aalph2}.
\subsubsection{Coordination with 2 cells}
The average cell-edge spectral efficiency for the proposed adaptive bit allocation scheme is shown in Fig.~\ref{Fig4} with $B_{\textrm{total}}=8$, $M=4$ and $L=2$. It is compared with the coordinated ZF adaptive bit allocation scheme \cite{5648782}. It is seen that the proposed scheme improves the average cell-edge spectral efficiency compared to \cite{5648782}. 
\begin{table} [!t]
\centering
    \caption{Average spectral efficiency performance with instantaneous schemes.}
    \label{tt1}
    \resizebox{9cm}{1cm}{
    \begin{tabular}{|l|c|c|c|c|}
  \hline
$\rho_{0}$        & -4 dB     & 0 dB& 2 dB& 6 dB\\
  \hline \hline
Maximizing inst. & 3.7  bps/Hz  & 5.5 bps/Hz& 6.2 bps/Hz&  7.5 bps/Hz \\
spectral efficiency   &  & &  &  \\
  \hline
Minimizing inst. & 3.5 bps/Hz& 5.4 bps/Hz& 6.2 bps/Hz& 7.3 bps/Hz \\
interference  & &  &   &  \\
  \hline
    \end{tabular}
    }
  \end{table}
In Table \ref{tt1}, we compare the average spectral efficiency performance with two schemes: a) maximizing instantaneous spectral efficiency and b) minimizing instantaneous interference. The performance of both schemes are nearly equivalent.
 \begin{figure}[h]
 \centering
 \includegraphics[width=.5\textwidth]{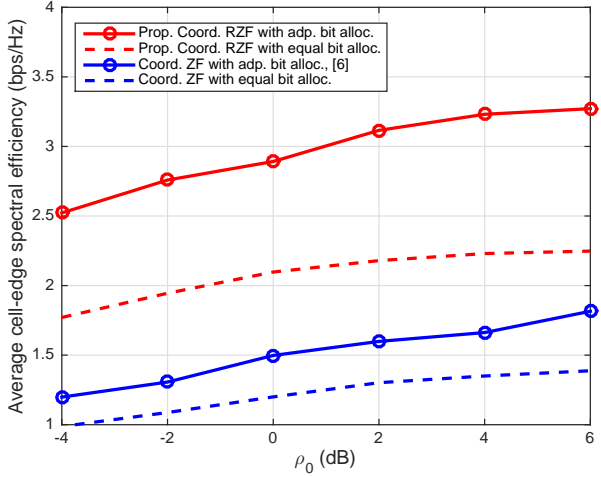}
 \caption{The average cell-edge spectral efficiency with $K=3$, $L=2$, $M=6$ and $B_{\textrm{total}}=9$.}
 \label{Fig6}
\end{figure}
\subsubsection{Coordination with 3 cells}
The average cell-edge spectral efficiency for $K=3$ cells is shown in Fig. \ref{Fig6} using the proposed coordinated RZF with adaptive bit allocation strategy, where $B_{\textrm{total}} = 9$, $M=6$ and $L=2$.
The proposed adaptive bit allocation strategy yields better cell-edge spectral efficiency compared to \cite{5648782}.

\section{Conclusion}
We analyzed a coordinated RZF precoding strategy for multicell MU MISO systems. We proposed an adaptive feedback bit allocation scheme with limited feedback RVQ CDI that minimizes the expected interference at users. The proposed adaptive bit allocation scheme yields higher cell-edge spectral efficiency than the existing coordinated ZF based adaptive bit allocation method.

\appendices

\section{Proof of Result 1}\label{aa1}
Here we provide the derivation details of Result 1. Let 
\begin{align}
 \text{D}^{(t)}_k &= \mathbb{E} \left[ \sum_{i=1}^M \frac{\left( \lambda_n \right)^t }{\left( \lambda_n + \alpha_k \right)^2}\right] 
  = \hspace{-.2em}M\hspace{-.5em}\int_0^{\infty}\hspace{-.8em}\frac{\left( \lambda \right)^t }{\left( \lambda + \alpha_k \right)^2}  f_{0} \left( \lambda \right)  d \lambda \label{a1}
\end{align}
where $\lambda$ is an arbitrary eigenvalue with probability density function (pdf) \cite{SmithS02}
\begin{equation}\label{a3}
 f_{0} \left(\lambda\right) = \frac{1}{M} \sum_{i=1}^M \text{e}^{-\lambda} \sum_{j=0}^{i-1} \left(-1\right)^{j} {i-1 \choose i-1-j} \frac{\lambda^j}{j!}
\end{equation}
Substituting \eqref{a3} in \eqref{a1} gives
\begin{flalign}
\text{D}^{(t)}_k &\hspace{-.5em}= \hspace{-.5em}\int_0^{\infty} \hspace{-1em}
 \frac{\left( \lambda \right)^t }{\left( \lambda + \alpha_k \right)^2} \sum_{i=1}^M \text{e}^{-\lambda}\hspace{-.4em}\left( \sum_{j=0}^{i-1} \left(-1\right)^{j} \hspace{-.4em}{i-1 \choose i-1-j} \frac{\lambda^j}{j!} \hspace{-.4em}\right)^2 \hspace{-.6em}d\lambda && \nonumber \\
 & = \sum_{i=1}^M \sum_{j=0}^{i-1} \sum_{l=0}^{i-1} (-1)^{j+l} {i-1 \choose i-1-j} {i-1 \choose i-1-l} \frac{1}{j!l!} && \nonumber \\
 & \ \ \ \times \int_{0}^{\infty} \frac{\lambda^{t+j+l} \text{e}^{-\lambda}}{\left( \lambda + \alpha_k\right)^2}  d\lambda.&& \label{ia11}
\end{flalign}
Substituting $\lambda= v-\alpha_k$ in the integral in \eqref{ia11} gives \eqref{fc6}.

\section{Proof of Result 2}\label{a2}
Let, $\text{F}_k$ be defined as
\begin{align}
 \text{F}_k &= \mathbb{E} \left[ \left( \sum_{n=1}^M \frac{\lambda_n}{\lambda_n + \alpha_k}\right)^2\right] \nonumber \\
 & = \text{D}^{(2)}_k  + \mathbb{E} \left[ \sum_{a=1}^M \sum_{b=1,b \neq a}^M \frac{\lambda_a}{\lambda_a + \alpha_k} \frac{\lambda_b}{\lambda_b + \alpha_k} \right], \label{b1}
\end{align}
where $\lambda_a$ and $\lambda_b$ are two distinct arbitrary eigenvalues. Denoting $f_{0} \left( \lambda_a,\lambda_b \right)$ as the joint pdf of two distinct arbitrary eigenvalues, we can write \eqref{b1} as
\begin{equation}\label{b2}
 \text{F}_k = \text{D}^{(2)}_k + M(M-1)\hspace{-.5em}  \int_{0}^{\infty}\hspace{-.5em} \int_{0}^{\infty} \hspace{-.8em} \frac{\lambda_a \lambda_b f_{0} \left(\lambda_a,\lambda_b \right) }{(\lambda_a +\alpha_k) (\lambda_b +\alpha_k)} d\lambda_a  d\lambda_b.
\end{equation}
The joint pdf of two distinct arbitrary eigenvalues is \cite{SmithS02}
\begin{equation}
 f_{0} \left( \lambda_a, \lambda_b \right) = \frac{1}{M(M-1)} \sum_{i=1}^{M} \sum_{j=1,j\neq i}^{M} \text{e}^{-(\lambda_a + \lambda_b)} \text{Z} (\lambda_a,\lambda_b),\nonumber
\end{equation}
where denoting $\textrm{L}_n(\cdot)$ as the $n^{\textrm{th}}$ Laguerre polynomial, we have
$\text{Z} (\lambda_a,\lambda_b) = \textrm{L}_{i-1} \left( \lambda_a \right)^2 \textrm{L}_{j-1} \left( \lambda_b \right)^2 - \textrm{L}_{i-1} \left( \lambda_a \right) \textrm{L}_{j-1} \left( \lambda_a \right) \textrm{L}_{i-1} \left( \lambda_b \right) \textrm{L}_{j-1} \left( \lambda_b \right)$. So now we can write \eqref{b2} as
\begin{align}
 \text{F}_k = &\text{D}^{(2)}_k + \sum_{i=1}^{M} \sum_{j=1,j\neq i}^{M}  \label{b3}\\
 &\underbrace{\int_{0}^{\infty} \hspace{-.7em} \int_{0}^{\infty} \hspace{-.7em} \frac{\lambda_a}{\lambda_a +\alpha_k}\frac{\lambda_b}{\lambda_b +\alpha_k} \text{e}^{-(\lambda_a + \lambda_b)} \text{Z} \left(\lambda_a,\lambda_b \right)  d\lambda_a  d\lambda_b}_{\textrm{Y}}.\nonumber
\end{align}
As the double integrals in \eqref{b3} are of the same function but with different variables, we can also write $Y$ as
\begin{flalign}
 \textrm{Y} &= \left(\int_{0}^{\infty}\frac{\text{e}^{-\lambda}\lambda}{\lambda +\alpha_k}  \textrm{L}_{i-1}  \left(\lambda \right)^2   d\lambda \right)^2 && \label{b5}\\
 & \ \ \ - \left(\int_{0}^{\infty}\frac{\text{e}^{-\lambda}\lambda}{\lambda +\alpha_k} \textrm{L}_{i-1} \left( \lambda \right) \textrm{L}_{j-1} \left( \lambda \right) d\lambda \right)^2 && \nonumber \\
  = &\left(\sum_{r=0}^{i-1} \sum_{s=0}^{i-1} \left( -1 \right)^{r+s} {i-1 \choose i-1-r}{i-1 \choose i-1-s} \frac{\Phi}{r! \ s!} \right)^2 && \nonumber \\
& -   \left(\sum_{r=0}^{i-1} \sum_{s=0}^{j-1} \left( -1 \right)^{r+s} {i-1 \choose i-1-r}{j-1 \choose j-1-s} \frac{\Phi}{r! \ s!}  \right)^2,&& \nonumber
\end{flalign}
where $ \Phi = \int_{0}^{\infty} \lambda^{1+r+s} \text{e}^{-\lambda}/\left( \lambda + \alpha_k\right)  d\lambda$. Solving the integrals in \eqref{b5} by substituting $v=\lambda+\alpha_k$ we get \eqref{n_eq1}.

\bibliographystyle{ieeetr}
\bibliography{MIRZA_VT-2015-01172.R2}

\begin{thebibliography}{10}

\bibitem{4487516}
J.~Zhang, R.~Chen, J.~Andrews, and R.~Heath, ``Coordinated multi-cell {MIMO}
  systems with cellular block diagonalization,'' in {\em Proc. Asilomar Conf.
  on Signal, Syst. and Comput.}, pp.~1669 -- 1673, 2007.

\bibitem{jindal2006mimo}
N.~Jindal, ``{MIMO} broadcast channels with finite-rate feedback,'' {\em IEEE
  Trans. Inf. Theory}, vol.~52, no.~11, pp.~5045 -- 5060, 2006.

\bibitem{bhagavatula2011adaptive}
R.~Bhagavatula and R.~W. Heath, ``Adaptive limited feedback for sum-rate
  maximizing beamforming in cooperative multicell systems,'' {\em IEEE Trans.
  Signal Process.}, vol.~59, no.~2, pp.~800 -- 811, 2011.

\bibitem{au2007performance}
C.~Au-Yeung and D.~Love, ``On the performance of random vector quantization
  limited feedback beamforming in a {MISO} system,'' {\em IEEE Trans. Wireless
  Commun.}, vol.~6, no.~2, pp.~458 -- 462, 2007.

\bibitem{zhang2010adaptive}
J.~Zhang and J.~G. Andrews, ``Adaptive spatial intercell interference
  cancellation in multicell wireless networks,'' {\em IEEE J. Sel. Areas
  Commun.}, vol.~28, no.~9, pp.~1455 -- 1468, 2010.

\bibitem{5648782}
N.~Lee and W.~Shin, ``Adaptive feedback scheme on {K-}cell {MISO} interfering
  broadcast channel with limited feedback,'' {\em IEEE Trans. on Wireless
  Commun.}, vol.~10, no.~2, pp.~401 -- 406, 2011.

\bibitem{1391204}
C.~Peel, B.~Hochwald, and A.~Swindlehurst, ``A vector-perturbation technique
  for near-capacity multiantenna multiuser communication-part {I:} {C}hannel
  inversion and regularization,'' {\em IEEE Trans. Commun.}, vol.~53, no.~1,
  pp.~195 -- 202, 2005.

\bibitem{HoydisBD13}
J.~Hoydis, S.~ten Brink, and M.~Debbah, ``Massive {MIMO} in the {UL/DL} of
  cellular networks: {H}ow many antennas do we need?,'' {\em {IEEE} J. Sel.
  Areas Commun.}, vol.~31, no.~2, pp.~160 -- 171, 2013.

\bibitem{6779600}
R.~Muharar, R.~Zakhour, and J.~Evans, ``Base station cooperation with feedback
  optimization: {A} large system analysis,'' {\em IEEE Trans. Inf. Theory},
  vol.~60, no.~6, pp.~3620 -- 3644, 2014.

\bibitem{6678041}
R.~Muharar, R.~Zakhour, and J.~Evans, ``Optimal power allocation and user
  loading for multiuser {MISO} channels with regularized channel inversion,''
  {\em IEEE Trans. Commun.}, vol.~61, no.~12, pp.~5030 -- 5041, 2013.

\bibitem{5510182}
L.~Yu, W.~Liu, and R.~Langley, ``{SINR} analysis of the subtraction-based {SMI}
  beamformer,'' {\em IEEE Trans. Signal Process.}, vol.~58, no.~11, pp.~5926 --
  5932, 2010.

\bibitem{4686268}
A.~Dabbagh and D.~Love, ``Multiple antenna {MMSE} based downlink precoding with
  quantized feedback or channel mismatch,'' {\em IEEE Trans. Commun.}, vol.~56,
  no.~11, pp.~1859 -- 1868, 2008.

\bibitem{6397539}
E.~Park, H.~Kim, H.~Park, and I.~Lee, ``Feedback bit allocation schemes for
  multi-user distributed antenna systems,'' {\em IEEE Commun. Lett.}, vol.~17,
  pp.~99 -- 102, 2013.

\bibitem{yu2012novel}
S.~Yu, H.-B. Kong, Y.-T. Kim, S.-H. Park, and I.~Lee, ``Novel feedback bit
  allocation methods for multi-cell joint processing systems,'' {\em IEEE
  Trans. Wireless Commun.}, vol.~11, no.~9, pp.~3030 -- 3036, 2012.

\bibitem{6585723}
J.~Zhang, C.-K. Wen, S.~Jin, X.~Gao, and K.-K. Wong, ``Large system analysis of
  cooperative multi-cell downlink transmission via regularized channel
  inversion with imperfect {CSIT},'' {\em IEEE Trans. Wireless Commun.},
  vol.~12, no.~10, pp.~4801 -- 4813, 2013.

\bibitem{nguyen2014mmse}
D.~H. Nguyen and T.~Le-Ngoc, ``{MMSE precoding for multiuser MISO downlink
  transmission with non-homogeneous user SNR conditions},'' {\em EURASIP J.
  Advances in Signal Process.}, no.~1, p.~85, 2014.

\bibitem{SmithS02}
P.~J. Smith and M.~Shafi, ``On a {G}aussian approximation to the capacity of
  wireless {MIMO} systems,'' in {\em Proc. IEEE Int. Conf. on Commun. (ICC)},
  pp.~406 -- 410, 2002.

\end{thebibliography}

\end{document}